\documentstyle[IJFCS]{article}

\begin{document}
%% print out the publisher copyright heading
\copyrightheading

%% use symbolic footnote
\symbolfootnote

%% use normal text like skip (13pt)
\textlineskip

\begin{center}

%% print out titles in IJFCS format
\fcstitle{The Existence of
$\omega$-Chains for
Transitive Mixed Linear\\
 Relations and Its Applications}

\vspace{24pt}

{\authorfont Zhe Dang\footnote{Corresponding author ({\tt zdang@eecs.wsu.edu}).}}

\vspace{2pt}

%% use smaller line skip here
\smalllineskip
{\addressfont School of Electrical Engineering and Computer Science,\\
Washington State University,
Pullman, Washington 99164, USA}

\vspace{10pt}
and

\vspace{10pt}
{\authorfont Oscar H. Ibarra}

\vspace{2pt}
\smalllineskip
{\addressfont Department of Computer Science,\\
University of California,
Santa Barbara, California 93106, USA}

\vspace{20pt}
%% authors need not care about this
\publisher{(received date)}{(revised date)}{Editor's name}

\end{center}

\alphfootnote

\begin{abstract}
We show that
it is decidable whether a transitive 
mixed linear relation has an $\omega$-chain.
Using this result,
we study
a number of liveness
verification problems for generalized timed automata
within a unified
framework.
More precisely, we prove that
(1) the mixed linear liveness
 problem for
a timed automaton with dense clocks, reversal-bounded
counters, and a free counter is decidable, and
(2) the Presburger liveness
 problem for
a timed automaton with discrete clocks, reversal-bounded
counters, and a pushdown stack is decidable.
\end{abstract}

%\footnotetext{This is an abstract footnote}

\textlineskip

\def\bfvec #1{{\mathbf{#1}}}
\renewcommand{\vec}{\bfvec}
\newcommand{\Z} {{\bf Z}}
\newcommand{\R} {{\bf R}}
\newcommand{\Q} {{\bf Q}}
\newcommand{\D} {{{\bf D}^+}}
\newcommand{\F} {{{\bf D}}}
\newcommand{\M}{{\cal M}}
\newcommand{\W} {{\vec W}}

\newcommand{\T} {{\cal T}}

\newcommand{\deq} {{\sim}}
\newcommand{\rdeq} {{\approx}}
\def\relv #1{\widehat{#1}}

\def\up #1{{\lceil{#1}\rceil }}
\def\down #1{{\lfloor{#1}\rfloor }}
\newcommand{\srto}{\leadsto^{\cal A}_{\langle s,X_r\rangle}}
\newcommand{\srleads}{\leadsto^{\cal A}_{X_r}}
\def\B{{\bf B}}
\def\A{{\cal A}}
\def\C{{\cal C}}
\def\P{{\cal P}}
\def\RPTA{{\cal B}}

\def \v{{\vec v}}
\def \x{{\vec x}}
\def \u{{\vec u}}
\def \X{{\vec X}}
\def \Y{{\vec Y}}
\def \Z{{\vec Z}}
\def \V{{\vec V}}
\def \U{{\vec U}}

\def \FNPCM {{(\bf \F+NPCA)}}
\def\ring #1{{\stackrel{\circ}{#1}}}
\def\qed{\hfill\rule{1.5mm}{1.5ex}}

\newcommand{\modd}{{\rm mod}}
\newcommand{\mod}{{~\equiv}}
\newcommand{\notmod}{{~\not\equiv}}

\def\integer #1{{#1}^{\bf Int}}
\def\dense #1{{#1}^{\bf Frac}}

\section{
Introduction}\label{chapt0}
In the area of model-checking, the search for efficient
techniques for verifying infinite-state systems has been
an ongoing research effort. Much work has been devoted
to investigating various restricted models of infinite-state
systems that are amenable to automatic verification
for some
 classes of temporal properties, e.g., safety and liveness.
A timed automaton is one such model.
%OO changed above
A timed automaton \cite{AD94}
is a finite automaton (over finitely many control states)
augmented with dense clocks.
The clocks
can be reset or progress at the same rate, and can be
tested against
clock constraints in the form of
clock regions
(i.e., comparisons of a clock or the difference of two clocks
against an integer constant, e.g., $x-y<6$, where $x$ and $y$ are clocks.).
 The most important
result in the theory of timed automata is that
region reachability for  timed automata
is decidable \cite{AD94}.
This result has been used in
defining various real-time logics,
%appropriate
%OO Delete above?
model checking
algorithms and tools
   \cite{ACD93,AH94,HH95,HNSY94,LPY97,LLW95,RS97,W94}
for verifying
real-time systems.

However, 
region reachability
 is  not strong enough 
to verify many complex timing requirements
not in the form of clock regions
(e.g., ``$x_1-x_2>2(x_3-x_4)$
 is always true") for timed automata.
Recently,  
decidable binary reachability
(i.e., the set of all
pairs of configurations such that one can reach the other)
 characterizations
for timed automata and their generalizations
were obtained \cite{CJ99,D01,DIBKS00}.
%OO were above
The characterizations
opened the door for automatic verification of
%OO changed line above
%OO opened above
various real-time models against complex timing requirements.
For instance, a flattening technique
was used by Comon and Jurski \cite{CJ99}
%OO was above
to establish that the binary reachability of
timed automata is definable in the additive theory of
the reals and integers. 
%OO put the above
A timed automaton can be
%further
%OO delete further
augmented with other
unbounded
discrete data structures such as
a free counter and
reversal-bounded counters.
A (free) counter
is an integer variable
that can be incremented by 1, decremented by 1, and
tested against 0.
A counter is reversal-bounded if
the number of times it alternates between nondecreasing
and nonincreasing mode and vice-versa
is bounded by some fixed number independent of the computation \cite{I78}.
%OO Change above 4 lines
A pattern technique was proposed by
%OO was above
Dang \cite{D01} to obtain a decidable 
binary reachability characterization on
some ``storage-augmented" timed automata.
%OO changed above
For instance, suppose that $\A$
is a timed automaton
%OO change above line
(with dense clocks $x_1$ and $x_2$)
augmented with two reversal-bounded counters $y_1$ and $y_2$,
and a free counter $y_3$.
The result of Dang \cite{D01} 
implies that
the binary reachability of $\A$ is definable 
in the additive theory of the reals and integers.
%OO put the above.
Therefore, we can automatically
verify the following safety property,
which contains linear constraints on {\em both} dense variables and
unbounded discrete variables,
\begin{quote}
``Given two control states $s_1$ and $s_2$,
if $\A$ starts at $s_1$ in a configuration
satisfying $x_1-2x_2+y_1-2y_2+y_3>5$,
then whenever $\A$ reaches $s_2$, its
configuration must 
satisfy
$x_1+x_2<y_2-2y_3+2$."
\end{quote}
%OO change above quote

In contrast to safety properties,
liveness properties considered in this paper involve 
properties on infinite
executions of $\A$.  
For instance, 
consider an
infinite
execution that
passes some control state for infinitely many times.
A mixed linear constraint on 
clocks and counters in $\A$ may or may not be satisfied
whenever $\A$ passes the control state.
Is there an infinite execution on which
the constraint is satisfied for infinitely many times
at the control state?
An example liveness  property
would be like below:
\begin{quote}
``Given two control states $s_1$ and $s_2$,
if $\A$ starts at $s_1$ in some configuration 
satisfying $x_1-2x_2+y_1-2y_2+y_3>5$,
then $\A$ has an infinite execution on which
$x_1+x_2<y_2-2y_3+2$ is satisfied at $s_2$ for infinitely many times."
\end{quote}
%OO changed above.  BUT it's not clear what the last line mean.
%%DD changed before the property statement.
This kind of liveness properties have
a lot of applications such as
whether
concurrent real-time processes are livelock-free, starvation-free, etc.
Can this liveness property be automatically verified for $\A$?

We approach this question
by looking at mixed linear relations $R$
 that are 
relations on real and integer variables
definable in the additive theory of the reals and integers.
%OO put the above
We first prove the main theorem that the existence of
an $\omega$-chain for $R$ is decidable when $R$ is
transitive. This proof is done by eliminating quantifiers
from $R$ using a recent result of 
\cite{Weis99} and expressing $R$ into 
mixed linear constraints.
The decidable result follows from the fact that
the existence of
an $\omega$-chain for $R$ forces $R$ to have a special format.
Notice that the transitivity of $R$ is critical;
removing it from $R$ obviously causes the existence of
an $\omega$-chain undecidable (e.g., encoding 
the one-step
transition relations of a two-counter machine into $R$).
%OO Is above obvious?
%DD add "one-step"

Recall that the binary reachability of $\A$ is a
transitive mixed linear relation.
The above liveness question can be reduced to the existence of
an $\omega$-chain for some mixed linear relation
easily constructed from the binary reachability.
Therefore, a direct application of the main theorem gives
a positive answer to the question.
%OO changed to positive
We may also use the main theorem to 
verify a class of pushdown systems. For instance, suppose
that $\cal P$ is a pushdown 
automaton.
Consider the following Presburger liveness  property:
\begin{quote}
``Given two states $s_1$ and $s_2$,
from some configuration at $s_1$ satisfying $n_a-2n_b>n_c$,
$\cal P$ has an infinite execution on which
$n_a+n_b<3n_c$
holds at $s_2$ for infinitely many times,"
\end{quote}
where count variable
$n_a$ indicates the number of 
symbol $a$'s in the stack word in  a configuration.
This paper provides a technique to reduce this property
into
 the existence of an 
$\omega$-chain for some Presburger relation, which
is a special form
of mixed linear relations.
Therefore, using the main theorem,
the above property can be automatically
verified for $\cal P$. In fact, 
we show the result for a more powerful
class of pushdown systems: $\cal P$ can be
a pushdown automaton augmented with reversal-bounded counters and
integer-valued clocks. This class
of pushdown systems can be used to model
a class of real-time recursive programs. The Presburger liveness
properties for
this class of pushdown systems
then  contain Presburger
formulas on
count variables, reversal-bounded counters and
discrete clocks.

The techniques presented in this paper are
%significantly
%OO delete above
different from 
our previous papers \cite{DSK01,DIS01}
on liveness verification.
In those two papers, we only deal with
the Presburger liveness problems
for discrete timed automata (i.e., timed automata with
integer-valued clocks) \cite{DSK01}
 and for 
reversal-bounded counter machines
with a free counter (NCMFs) \cite{DIS01},
respectively.
Both of the papers are based upon analyzing
loops in 
the machines. In particular,
the key idea in \cite{DSK01} is to make discrete timed automata
static (i.e., enabling conditions can be removed)
and memoryless (i.e., two integer clock values are somewhat unrelated if
they are separated by a large number of clock resets).
But, the idea cannot be easily extended to dense clocks.
The key idea in \cite{DIS01}
is to partition an execution of an NCMF into phases
such that reversal-bounded counters are monotonic in each phase.
Then, a technique is used to reduce 
the NCMF into one with only one free counter, with respect to
the liveness property. But, we were not able to 
extend the idea when the free counter is replaced by a pushdown stack.
The techniques presented in this paper, however, 
allows us to handle,
in a unified framework,
 a stronger class of systems:
timed automata with dense clocks,
reversal-bounded counters, and
a free counter. In addition, we can deal with a class of
generalized pushdown systems.

The paper is organized as follows.
Section \ref{chaptPrelimi}
 gives the basic definitions and preliminary results
that are used in the paper.
Sections \ref{s3} through \ref{s5} present
the proof of the main theorem; i.e.,
it is decidable whether a transitive 
mixed linear relation has an $\omega$-chain.
Section \ref{s6} applies the main theorem in showing
the decidable results on
the mixed linear liveness problem for
a timed automaton augmented with reversal-bounded counters and
a free counter and on
the Presburger liveness problem for
a discrete timed automaton
augmented with reversal-bounded counters and
a pushdown stack.
Finally, Section \ref{s7} concludes with some
remarks.
%OO Changed two lines above

\section{
Preliminaries}\label{chaptPrelimi}

Let $m$ and $n$ be positive integers.
Consider a formula
$$\sum_{1\le i\le m} a_ix_i+
\sum_{1\le j\le n} b_jy_j \sim c,$$
where each
$x_i$ is a real variable,
each $y_j$ is an integer variable,
each $a_i$, each $b_j$ and $c$
are integers,
$1\le i\le m, 1\le j\le n$,
and $\sim$ is
$=$, $>$, or
$\equiv_d$ for some integer $d>0$.
The formula is a
 {\em
mixed linear constraint}
if $\sim$ is
$=$ or $>$.
The formula is called 
a {\em dense linear constraint}
if $\sim$ is
$=$ or $>$ and
each $b_j=0$, $1\le j\le n$.
The formula is called
a {\em discrete  linear constraint}
if 
$\sim$ is
 $>$ and
each $a_i=0$, $1\le i\le m$.
The formula is called
a {\em
discrete mod constraint}, if
each $a_i=0$, $1\le i\le m$,
and $\sim$ is $\equiv_d$ for some integer $d>0$.

A formula is {\em definable in the
additive theory of reals and integers 
(resp.  reals, integers)} if it is the result of
%OO Changed above line
applying quantification ($\exists$)
 and Boolean operations ($\neg$ and $\land$)
over mixed linear constraints
(resp. dense linear constraints,
discrete linear constraints);
the formula is called a {\em mixed formula
(resp. dense formula, Presburger formula)}.
It is decidable whether the formula is satisfiable.
It is well-known that a Presburger formula 
 can always be written, after quantifier elimination,
 as a disjunctive normal
form of 
discrete linear constraints
and discrete mod constraints.
It is also known that
a dense formula can always be written as a disjunctive normal
form of 
dense linear constraints.
Can we eliminate quantifiers in  mixed formulas?
%DD change below
The answer is not obvious. This is because
a mixed formula like $\exists y (x_1-x_2=y)$,
after eliminating all the quantifiers,
is not always in the form of a Boolean combination of
mixed linear constraints.
%OO What is the purpose of the above sentence?  Perhaps delete the sentence?
%DD changed

%OO Dang: The notation below x` and x' is not easy to read.
%% Perhaps use $x^{(i)}$ and $x^{(r)}$ or similar notation instead?
%% DD: done
A real variable $x$ can be treated as the sum of
an integer  variable (the integral part of $x$)
$\integer x$
and
a real variable (the fractional part of $x$)
$\dense x$
with $x=\integer{x}+\dense x$ and $0\le \dense x<1$.
A mixed formula $R(x_1,\cdots, x_m, y_1,\cdots, y_n)$,
where $x_1,\cdots, x_m,$ $y_1,\cdots, y_n$ are the free
variables,
can therefore  be translated into
another mixed formula $\hat R$ (called $R$'s {\em separation}):
$$R(\integer x_1+\dense x_1,
\cdots, \integer x_m+\dense x_m,
y_1,\cdots, y_n)\land 0\le \dense x_1<1
\land \cdots \land 0\le \dense x_m<1.$$
Notice that the separation $\hat R$ contains 
real variables $\dense x_1,\cdots,\dense x_m$ and
integer variables
$\integer x_1,\cdots, \integer x_m,$ $
y_1,\cdots, y_n$.
The following result can be easily obtained from
\cite{Weis99}, in which
the separation  can be written into
a Boolean combination of
dense linear constraints,
discrete linear constraints,
and discrete mod constraints.
A nice property of the Boolean combination
is that real variables 
and integer variables
 are separated:
each constraint in the combination
 either  contains 
real variables
$\dense x_1,\cdots,\dense x_m$
 only or 
contains 
integer variables
$\integer x_1,\cdots, \integer x_m,$ 
$y_1,\cdots, y_n$ only.

\begin{theorem}\label{Weis}
The separation of
any mixed formula
can be written into
a Boolean combination of
dense linear constraints,
discrete linear constraints,
and discrete mod constraints.
\end{theorem}

\begin{definition}
$R$ is a {\em mixed linear relation}
if it is a mixed formula  $R(\X,\Y,\X',\Y')$
over $2m$ real variables
$\X=x_1,\cdots,x_m {\rm ~and~} \X'=x_1',\cdots,x_m'$
and $2n$ integer variables
$\Y=y_1,\cdots,y_n {\rm ~and~} \Y'=y_1',\cdots,y_n'.$
\end{definition}

We use $\vec U$
%(with $\vec U_i$, $1\le i\le m$, being 
%the $i$-th component)
to denote an $m$-ary real vector
and use $\vec V$
to denote an $n$-ary integer vector.

\begin{definition}
A mixed linear relation $R$ is {\em transitive} if
for all $\vec U, \vec V, \vec U',\vec V',\vec U'',\vec V''$,
$R(\vec U, \vec V, \vec U',\vec V')\land
R(\vec U',\vec V',\vec U'',\vec V'')$
implies
$R(\vec U, \vec V,\vec U'',\vec V'')$.
An infinite sequence 
$(\vec U^0,\vec V^0),\cdots,
(\vec U^k,\vec V^k),\cdots$
is {\em an $\omega$-chain}
of  $R$ if
$R(\vec U^k,\vec V^k,\vec U^{k+1},\vec V^{k+1})$
holds for all $k\ge 0$.
The sequence is a {\em strong $\omega$-chain}
of  $R$ if
it is an $\omega$-chain of  $R$ 
satisfying
$R(\vec U^{k_1},\vec V^{k_1},\vec U^{k_2},\vec V^{k_2})$
for all $0\le k_1<k_2$.
\end{definition}

Notice that, if $R$ is transitive, then any 
subsequence
$$(\vec U^{i_0},\vec V^{i_0}),\cdots,
(\vec U^{i_k},\vec V^{i_k}),\cdots$$
(with $0\le i_0<\cdots < i_k<\cdots$)
of an $\omega$-chain $(\vec U^0,\vec V^0),\cdots,
(\vec U^k,\vec V^k),\cdots$
is also an
$\omega$-chain of $R$.
According to the definition of the separation $\hat R$
(which is also a mixed linear relation)
of  a mixed linear relation $R$ and Theorem \ref{Weis},
the following lemma can be proved.

\begin{lemma}\label{l1}
(1). A mixed linear relation is transitive iff
its separation is transitive.
(2). A mixed linear relation has an $\omega$-chain
iff its separation has an $\omega$-chain.
\end{lemma}

\section{A Technical Lemma}\label{s3}

We will  show that it is decidable whether
a transitive mixed linear relation $R$
has an $\omega$-chain.
 From Lemma \ref{l1},
it suffices  to work on the 
separation of $R$; i.e., from Theorem \ref{Weis},
we assume that $R$ itself is already in the form of
a Boolean combination of
dense linear constraints
(with each real variable taking values in $[0,1)$),
discrete linear constraints,
and discrete mod constraints.
That is,
$R(\X,\Y,\X',\Y')$
can be written as a disjunction
\begin{equation}\label{eq00}
R_1\lor\cdots\lor R_p
\end{equation}
for some $p$, where
each $R_i$ is a conjunction of
$$S_i\land T_i.$$
Each $S_i$ is a conjunction of
$l$ dense linear equations
\begin{equation}\label{eq1}
\bigwedge_{1\le j\le l} P_{ij}^1(\X)+Q_{ij}^1(\X')=c_{ij}^1,
\end{equation}
followed by $l$ dense linear inequalities
\begin{equation}\label{eq2}
\bigwedge_{1\le j\le l} P_{ij}^2(\X)+Q_{ij}^2(\X')>c_{ij}^2,
\end{equation}
with $\X$ and $\X'$ taking values in
$[0,1)^m$.
Each $T_i$ is 
a conjunction of
$l$ discrete linear inequalities
\begin{equation}\label{eq3}
\bigwedge_{1\le j\le l} P_{ij}^3(\Y)+Q_{ij}^3(\Y')>c_{ij}^3,
\end{equation}
followed by $l$ discrete mod constraints
\begin{equation}\label{eq4}
\bigwedge_{1\le j\le l} P_{ij}^4(\Y)+Q_{ij}^4(\Y')\mod_{d_{ij}}~c_{ij}^4.
\end{equation}
Notice that
discrete linear equations like $y_1+2y_2=3$ can be expressed in
discrete linear inequalities
such as
$y_1+2y_2>2\land -y_1-2y_2>-4$.
Also notice that the negation of 
a discrete mod constraint like
$y_1+2y_2\notmod_{5}~3$ can be expressed into
a finite disjunction of mod constraints in (\ref{eq4}).
Each $P_{ij}^{h}$ and each
$Q_{ij}^h$ for $h=1,2$ (resp. $h=3,4$) are linear combinations
(with integer coefficients) over
real variables (resp. integer variables).

Mod constraints in (\ref{eq4})
can be eliminated using the following procedure.
Take $$d=\prod_{1\le i\le p, 1\le j\le l} d_{ij}.$$
Let $\vec d$ be an $n$-ary integer vector 
taking values in $\{0,\cdots,d-1\}^n$.
Let $R'(\X,\Z,\X',\Z')$  be
$$\bigvee_{\vec d,\vec d'} R(\X, d\Z+\vec d,\X',d\Z'+\vec d')$$
by substituting $\Y$ with $d\Z+\vec d$ and
$\Y'$ with $d\Z'+\vec d'$ in
$R(\X,\Y,\X',\Y')$, for all possible choices  of 
$\vec d$ and $\vec d'$. Clearly,
\begin{itemize}
\item $R$ is transitive iff $R'$ is transitive, and
\item $R$ has an $\omega$-chain iff
$R'$ has an $\omega$-chain.
\end{itemize}
In $R'$, there are no mod-constraints, since,
after the substitution,
%substituting $\Y$ and $\Y'$ with $\Z$ and $\Z'$,
the truth value of each mod-constraint in (\ref{eq4})
is known (according to the choice of $\vec d$ and $\vec d'$).
Hence,  we may assume that
%OO Changed above line
$R$ itself does not contain mod-constraints in  (\ref{eq4}). 

Consider an infinite sequence $\cal C^\omega$
$$(\vec U^0,\vec V^0),\cdots,
(\vec U^k,\vec V^k),\cdots.$$
Let $f(\X,\Y)$ be a term that is a linear combination of
real variables $\X$ and integer variables $\Y$.
The term  is {\em increasing}
(resp. {\em decreasing, flat}) on $\cal C^\omega$
if
$f(\U^k,\vec V^k)<f(\U^{k+1},\vec V^{k+1})$ (resp. $f(\U^k,\vec V^k)>
f(\U^{k+1},\vec V^{k+1})$,
$f(\U^k,\vec V^k)=f(\U^{k+1},\vec V^{k+1})$),
 for each $k\ge 0$.
The term   is {\em bounded increasing}
(resp. {\em bounded decreasing}) on $\cal C^\omega$
if $f$ is increasing (resp. decreasing)
on $\cal C^\omega$ and
there is a number $b$ such that
$f(\U^k,\vec V^k)<b$ (resp. $f(\U^k,\vec V^k)>b$)
for all $k\ge 0$.
The term   is {\em unbounded increasing}
(resp. {\em unbounded decreasing}) on $\cal C^\omega$
if $f$ is increasing (resp. decreasing)
on $\cal C^\omega$ and
$f$ is not bounded increasing (resp. decreasing)
on $\cal C^\omega$.
The term of $f$ could (but need not) be in one of 
the following five {\em modes}
on $\cal C^\omega$:

(mode1) unbounded increasing,

(mode2) unbounded decreasing,

(mode3) flat,

(mode4) bounded increasing,

(mode5) bounded decreasing.

\noindent Clearly, when $f$ only contains real variables,
(mode1) and (mode2) are impossible
(since each real variables is assumed in
$[0,1)$);
when $f$ only contains integer variables,
(mode4) and (mode5) are impossible.

We observe that, since $R$ is transitive,
$R$ has an $\omega$-chain iff
$R$ has an $\omega$-chain $\cal C^\omega$ on which
 each real variable $x\in \X$
(as well as
each integer variable $y\in \Y$, and
each term $P_{ij}^h$ and $Q_{ij}^h$, $h=1,2,3$,
$1\le i\le p,1\le j\le l$)
is in one of the five modes on $\cal C^\omega$.
A {\em mode vector} $\M$ is used to
indicate the chosen mode for each of the
variables and the terms.
There are at most
$3^m3^n3^{3pl}3^{3pl}$ distinct mode vectors.
Therefore, in order to decide
whether
$R$ has an $\omega$-chain,
we only need to decide
whether
$R$ has an $\omega$-chain with
 some
 mode
vector $\M$. In the sequel, we use the following abbreviation.

\begin{definition}
An $\omega$-chain is {\em monotonic of mode $\M$}
(or simply, {\em monotonic} when $\M$ is understood)
if the chain is with mode
vector $\M$. 
\end{definition}

Now, we are ready to prove the following lemma
using the pigeon-hole principle.

\begin{lemma}\label{l2}
Suppose that $R$ is a transitive
mixed linear relation in the form of
$R=R_1\lor\cdots\lor R_p$
where each $R_i$ is a conjunction of
atomic formulas in (\ref{eq1},\ref{eq2},\ref{eq3},\ref{eq4}).
Then,
$R$ has an $\omega$-chain iff
$R_i$ has a monotonic and strong
$\omega$-chain for some $1\le i\le p$
and some mode vector $\M$.
\end{lemma}

\proof{
($\Rightarrow$).
Assume that $R$ has 
an $\omega$-chain $\cal C^\omega$
\begin{equation}\label{eq5}
(\vec U^0,\vec V^0),\cdots,
(\vec U^k,\vec V^k),\cdots
\end{equation}
that is monotonic for some mode vector $\M$.
$R(\vec U^{k_1},\vec V^{k_1},\vec U^{k_2},\vec V^{k_2})$
holds for any $0\le k_1<k_2$, since
$R$ is transitive.
Recall that $R=R_1\lor\cdots \lor R_p$.
Notice that each $R_i$ is not necessarily transitive.
The following technique generalizes the one presented in \cite{DIS01}.
We use a predicate $I(k_1,k_2,i)$ to indicate
$0\le k_1<k_2\land R_i(\vec U^{k_1},\vec V^{k_1},\vec U^{k_2},\vec V^{k_2})$.
Clearly, for any
$k_1,k_2$ with
$0\le k_1<k_2$, there is an $i$ ($1\le i\le p$)
such that
$I(k_1,k_2,i)$
holds. Define $I'(k_1,i)$ as
$\forall k\exists k_2 (k_2>k \land I(k_1,k_2,i)).$
Hence, $I'(k_1,i)$ is true iff
there are infinitely many $k_2$ satisfying $I(k_1,k_2,i)$.
Since $i$ is bounded (i.e., $1\le i\le p$),
for each $k_1$, there is an $i$ satisfying
$I'(k_1,i)$.
Therefore, there is an $i_0$ ($1\le i_0\le p$),
such that
\begin{equation}\label{eq6}
\forall k\exists k_1 (k_1>k \land I'(k_1,i_0)).
\end{equation}
That is,
there are  infinitely many $k_1$ satisfying $I'(k_1,i_0)$.
According to the definition of
$I'$ and $I$,
formula
(\ref{eq6})
can be translated back to the following formula:
\begin{equation}\label{eq7}
\forall k\exists k_1>k
\forall k'>k_1\exists k_2>k'
R_{i_0}(\vec U^{k_1},\vec V^{k_1},\vec U^{k_2},\vec V^{k_2}).
\end{equation}
Since $\cal C^\omega$ is monotonic,
there is a $\U\in [0,1]^m$ such that
$\lim\U^k=\U$.
In addition,
$Q_{i_0j}^1(\U^k)$,
$Q_{i_0j}^2(\U^k)$,
and
$Q_{i_0j}^3(\V^k)$
in $R_{i_0}$ ($R_{i_0}$ is given 
in the form of  (\ref{eq1}),(\ref{eq2}), and (\ref{eq3}))
are all monotonic wrt $k$.
Hence, formula (\ref{eq7})
can be strengthened
into
\begin{equation}\label{eq8}
\forall k\exists k_1>k\exists k'>k_1\forall k_2>k'
R_{i_0}(\vec U^{k_1},\vec V^{k_1},\vec U^{k_2},\vec V^{k_2}).
\end{equation}
That is, there are infinitely many $k_1$ such that,
for each of these $k_1$,
there is a $k'>k_1$ satisfying
$R_{i_0}(\vec U^{k_1},\vec V^{k_1},\vec U^{k_2},\vec V^{k_2})$
for all $k_2>k'$.
 From these infinitely many $k_1$'s,
we select any strictly increasing infinite sequence
$$k_1^0,\cdots,k_1^q,\cdots.$$
For each $k_1^q$,
we can pick a $k_2^q$ from (\ref{eq8})
(treating
$k_1^q$ as $k_1$ and
$k_2^q$ as $k_2$).
By making each $k_2^q$ large enough,
we can obtain a strictly increasing infinite sequence
$$k_2^0,\cdots,k_2^q,\cdots.$$
Notice that, from (\ref{eq8}),
for each $q$,
\begin{equation}\label{eq9}
\forall k\ge k_2^q
R_{i_0}(\vec U^{k_1^q},\vec V^{k_1^q},\vec U^{k},\vec V^{k}).
\end{equation}
Now, we define a sequence
of indices as follows.
Let $t_0=0$.
Pick $t_1$ as any number satisfying
$t_0<t_1$ and $k_2^{t_0}<k_1^{t_1}$.
Pick $t_2$ as any number satisfying
$t_1<t_2$ and $k_2^{t_1}<k_1^{t_2}$, and so on.
The existence of 
each $t_q$ is guaranteed by 
the monotonicity of the two sequences
$k_1^0,\cdots,k_1^q,\cdots$
and $k_2^0,\cdots,k_2^q,\cdots.$
It is easy to verify
$$R_{i_0}(\vec U^{k_1^{t_q}},\vec V^{k_1^{t_q}},\vec U^{k_1^{t_{q+1}}},
\vec V^{k_1^{t_{q+1}}})$$
holds for each $q\ge 0$ according to the choice
of each $t_q$ and (\ref{eq9}).
Hence,
$$(\U^{k_1^{t_0}},\vec V^{k_1^{t_0}}),
\cdots,
(\vec U^{k_1^{t_q}},\vec V^{k_1^{t_q}}),\cdots$$
is an 
$\omega$-chain of $R_{i_0}$, which is also 
monotonic of mode $\M$. Notice that the $\omega$-chain
is also a strong $\omega$-chain of $R_{i_0}$. This is because
of the definition of $t_q$ and (\ref{eq9}). 
Therefore, we have already shown that,
if $R$ has an $\omega$-chain, then
$R_{i_0}$ has a monotonic and strong
$\omega$-chain for some $i_0$
and $\M$.

($\Leftarrow$). Obvious. }%end of proof

Recall that $R_i=S_i\land T_i$
where $S_i$ contains only dense variables
and $T_i$ contains only integer variables.
Therefore, for any $\M$,
$R_i$  has a monotonic and strong
$\omega$-chain  iff
both $S_i$ and $T_i$ have a monotonic and strong
$\omega$-chain.
Hence, from now on, we will focus on
$S_i$ and $T_i$ separately by looking at the following two problems:

1. whether $S$ has a monotonic and strong
$\omega$-chain, where $S$
is a conjunction of
dense linear equations in (\ref{eq1}) and
inequalities in (\ref{eq2});

2. whether $T$ has a monotonic and strong
$\omega$-chain, where $T$
is a conjunction of
integer linear inequalities in (\ref{eq3}).

Notice that $S$ and $T$ are not 
necessarily transitive.
Solutions to the  problems 
are given in the following two sections.

\section{The Existence of $\omega$-chains for
Dense Linear Equations and Inequalities}\label{s4}

Assume that
$S$ is a conjunction of
$l$ 
dense linear equations
$P_{j}^1(\X)+Q_{j}^1(\X')=c_{j}^1$
and
$l$ dense linear inequalities
$P_{j}^2(\X)+Q_{j}^2(\X')>c_{j}^2$.
Each dense variable takes values in $[0,1)$.
Let $\M$ be a mode vector 
(on each dense variable, each term $P_{j}^1$,
$Q_{j}^1$, $P_{j}^2$, $Q_{j}^2$, $1\le j\le l$).
We use ``$\nearrow$",
``$\to$" and
``$\searrow$" to stand for 
``bounded increasing", ``flat"
and ``bounded decreasing", respectively
(the other two modes ``unbounded increasing"
and ``unbounded decreasing" are not possible for 
dense variables and dense terms).
Assume that
$$\U^0,\cdots,\U^k,\cdots$$
is a monotonic and strong
$\omega$-chain $\U^\omega$
of $S$, for a given $\M$.
Therefore,
$S(\U^{k_1},\U^{k_2})$ holds
for any $0\le k_1<k_2$
(notice that
$S$ itself is not necessarily transitive.).
Since dense variables take values in $[0,1)$,
we have $\lim \U^k=\U$ for some $\U\in [0,1]^m$.

A number of observations can be made on $\U^\omega$ and $\M$.
For instance, each variable $x\in \X$
(as well
as each term $P_{j}^1$, $Q_{j}^1$,
$P_{j}^2$, $Q_{j}^2$) has a mode
 (given in $\M$)
 on $\U^\omega$.
In particular,
for a linear equation like
$P_{j}^1(\X)+Q_{j}^1(\X')=c_{j}^1$,
the mode of $P_{j}^1$ and the mode
of $Q_{j}^1$ must be flat.
How about
a linear inequality like 
$P_{j}^2(\X)+Q_{j}^2(\X')>c_{j}^2$?
Let us consider the case when $\M(P_{j}^2)=\searrow$
and $\M(Q_{j}^2)=\nearrow$.
In this case, since 
$\lim \U^k=\U$, we can easily conclude that, for any $k_1<k_2$,
$P_{j}^2(\U^{k_1})>P_{j}^2(\U^{k_2})>P_{j}^2(\U)$,
$Q_{j}^2(\U^{k_1})<Q_{j}^2(\U^{k_2})<Q_{j}^2(\U)$,
$P_{j}^2(\U)+Q_{j}^2(\U)\ge c_{j}^2$.
Similar conclusions can be made for
all the other possible
choices for  $\M(P_{j}^2)$ and $\M(Q_{j}^2)$.
Combining all these observations,
we obtain that,
for any $k_1<k_2$,
$H(\U,\U^{k_1},\U^{k_2},\M)$ holds, where
$H$ is defined as follows:
\begin{itemize}
\item $\U^{k_1}$ and $\U^{k_2}$ are consistent to the
mode $\M(x)$ for each $x\in \X$. That is, for all $x\in \X$,
$\U^{k_1}(x)<\U^{k_2}(x)$ (resp. $=$, $>$)
and $\U^{k_2}(x) \le \U(x)$ (resp. $=$, $\ge$) if
$\M(x)=\nearrow$ (resp. $\to$, $\searrow$), where
$\U^{k_1}(x)$ is the component for 
variable $x$ in vector $\U^{k_1}$.
\item  For each  
linear equation $P_{j}^1(\X)+Q_{j}^1(\X')=c_{j}^1$,
both $\M(P_{j}^1)$ and $\M(Q_{j}^1)$ must be
flat.
In this case,
$P_{j}^1(\U)+Q_{j}^1(\U)=c_{j}^1$,
$P_{j}^1(\U^{k_1})=P_{j}^1(\U^{k_2})=P_{j}^1(\U)$,
$Q_{j}^1(\U^{k_1})=Q_{j}^1(\U^{k_2})=Q_{j}^1(\U)$.
\item For each
linear inequality
$P_{j}^2(\X)+Q_{j}^2(\X')>c_{j}^2$,
according to each
possible combination
of $\M(P_{j}^2)$ and $\M(Q_{j}^2)$,
one of the following nine cases is satisfied:
\begin{itemize} 
\item $\M(P_{j}^2)=\nearrow$
and $\M(Q_{j}^2)=\nearrow$.
$P_{j}^2(\U^{k_1})<P_{j}^2(\U^{k_2})<P_{j}^2(\U)$,
$Q_{j}^2(\U^{k_1})<Q_{j}^2(\U^{k_2})<Q_{j}^2(\U)$,
and $P_{j}^2(\U)+Q_{j}^2(\U)>c_{j}^2$,
\item $\M(P_{j}^2)=\nearrow$
and $\M(Q_{j}^2)=\to$.
$P_{j}^2(\U^{k_1})<P_{j}^2(\U^{k_2})<P_{j}^2(\U)$,
$Q_{j}^2(\U^{k_1})=Q_{j}^2(\U^{k_2})=Q_{j}^2(\U)$,
$P_{j}^2(\U)+Q_{j}^2(\U)>c_{j}^2$,
\item $\M(P_{j}^2)=\nearrow$
and $\M(Q_{j}^2)=\searrow$.
$P_{j}^2(\U^{k_1})<P_{j}^2(\U^{k_2})<P_{j}^2(\U)$,
$Q_{j}^2(\U^{k_1})>Q_{j}^2(\U^{k_2})>Q_{j}^2(\U)$,
$P_{j}^2(\U)+Q_{j}^2(\U)>c_{j}^2$,
\item $\M(P_{j}^2)=\to$
and $\M(Q_{j}^2)=\nearrow$.
$P_{j}^2(\U^{k_1})=P_{j}^2(\U^{k_2})=P_{j}^2(\U)$,
$Q_{j}^2(\U^{k_1})<Q_{j}^2(\U^{k_2})<Q_{j}^2(\U)$,
$P_{j}^2(\U)+Q_{j}^2(\U)>c_{j}^2$,
\item $\M(P_{j}^2)=\to$
and $\M(Q_{j}^2)=\to$.
$P_{j}^2(\U^{k_1})=P_{j}^2(\U^{k_2})=P_{j}^2(\U)$,
$Q_{j}^2(\U^{k_1})=Q_{j}^2(\U^{k_2})=Q_{j}^2(\U)$,
$P_{j}^2(\U)+Q_{j}^2(\U)>c_{j}^2$,
\item $\M(P_{j}^2)=\to$
and $\M(Q_{j}^2)=\searrow$.
$P_{j}^2(\U^{k_1})=P_{j}^2(\U^{k_2})=P_{j}^2(\U)$,
$Q_{j}^2(\U^{k_1})>Q_{j}^2(\U^{k_2})>Q_{j}^2(\U)$,
$P_{j}^2(\U)+Q_{j}^2(\U)\ge c_{j}^2$,
\item $\M(P_{j}^2)=\searrow$
and $\M(Q_{j}^2)=\nearrow$.
$P_{j}^2(\U^{k_1})>P_{j}^2(\U^{k_2})>P_{j}^2(\U)$,
$Q_{j}^2(\U^{k_1})<Q_{j}^2(\U^{k_2})<Q_{j}^2(\U)$,
$P_{j}^2(\U)+Q_{j}^2(\U)\ge c_{j}^2$,
\item $\M(P_{j}^2)=\searrow$
and $\M(Q_{j}^2)=\to$.
$P_{j}^2(\U^{k_1})>P_{j}^2(\U^{k_2})>P_{j}^2(\U)$,
$Q_{j}^2(\U^{k_1})=Q_{j}^2(\U^{k_2})=Q_{j}^2(\U)$,
$P_{j}^2(\U)+Q_{j}^2(\U)\ge c_{j}^2$,
\item $\M(P_{j}^2)=\searrow$
and $\M(Q_{j}^2)=\searrow$.
$P_{j}^2(\U^{k_1})>P_{j}^2(\U^{k_2})>P_{j}^2(\U)$,
$Q_{j}^2(\U^{k_1})>Q_{j}^2(\U^{k_2})>Q_{j}^2(\U)$,
$P_{j}^2(\U)+Q_{j}^2(\U)\ge c_{j}^2$.
\end{itemize} 
\end{itemize}
Since $\lim \U^k=\U$, we have
$$\forall \delta>0 \exists
\U'\in [0,1)^m\forall \delta'>0\exists \U''\in[0,1)^m$$
\begin{equation}\label{eq10}
%\forall \delta>0 \exists 
%\U'\in [0,1)^m\forall \delta'>0\exists \U''\in[0,1)^m\linebreak
(H(\U,\U',\U'',\M)\land
|\U'-\U|<\delta \land
|\U''-\U|<\delta')
\end{equation}

Conversely, we can show  the following lemma.

\begin{lemma}
If there are a $\U\in [0,1]^m$ and
a mode vector $\M$ satisfying formula (\ref{eq10}),
then 
$S$ has a monotonic (of mode $\M$) and
strong $\omega$-chain in $[0,1)^m$.
\end{lemma}

\proof{
Assume (\ref{eq10}) holds for some
$\U\in [0,1]^m$ and
a mode vector $\M$. That is,
we can pick a sequence in $[0,1)^m$
$$\W^0,\cdots,\W^k,\cdots$$
such that,
\begin{itemize}
%\item Each $x\in \X$ has mode $\M(x)$ on the sequence,
\item $\lim \W^k=\U$,
\item $H(\U,\W^0,\W^k,\M)$ for each $k\ge 1$.
\end{itemize}
According to the fact that
$\lim \W^k=\U$ and 
the first item in the definition of
$H$, we can always pick a subsequence of
$\W^0,\cdots,\W^k,\cdots$
such that
each $x\in \X$ has mode $\M(x)$ on the subsequence.
Without loss of generality, we assume
that $\W^0,\cdots,\W^k,\cdots$
itself is the subsequence.

 From the definition of $H$,
for each
linear equation $P_{j}^1(\X)+Q_{j}^1(\X')=c_{j}^1$,
$\M(P_{j}^1)$ and $\M(Q_{j}^1)$ must both be flat.
In addition,
$P_{j}^1(\U)+Q_{j}^1(\U)=c_{j}^1$,
$P_{j}^1(\W^0)=P_{j}^1(\W^k)=P_{j}^1(\U)$,
$Q_{j}^1(\W^0)=Q_{j}^1(\W^k)=Q_{j}^1(\U)$.
Therefore, 
$\W^0,\cdots,\W^k,\cdots$
(as well as any subsequence)
 is already a strong $\omega$-chain for
the conjunction of these linear equations.
Clearly, each $P_{j}^1$ and
each $Q_{j}^1$ are in mode
$\M(P_{j}^1)=\M(Q_{j}^1)=\to$
on the chain.
In the rest of the proof, a ``subsequence"
always starts from $\W^0$.

For each
linear inequality
$P_{j}^2(\X)+Q_{j}^2(\X')>c_{j}^2$,
we will show that
a subsequence of
$\W^0,\cdots,\W^k,\cdots$
can be picked such that
the subsequence is a strong $\omega$-chain of the 
linear inequality, and
any subsequence
of the subsequence is also 
a strong $\omega$-chain of the 
linear inequality. In addition,
$P_{j}^2$ and
$Q_{j}^2$ are in modes $\M(P_{j}^2)$
and $\M(Q_{j}^2)$ on the subsequence, respectively.
By working on each linear inequality
one by one, a subsequence can be eventually picked
which is a monotonic (of mode $\M$) and
 strong $\omega$-chain of $S$.
Once this is done,
the lemma follows.

There are nine cases for the mode choices of
$\M(P_{j}^2)$ and $\M(Q_{j}^2)$.
We only prove the case when
$\M(P_{j}^2)=\searrow$
and $\M(Q_{j}^2)=\nearrow$;
all the other cases
can be shown analogously.
In the case, according to the definition of 
$H$, for each $k\ge 1$,
$P_{j}^2(\W^0)>P_{j}^2(\W^k)>P_{j}^2(\U)$,
$Q_{j}^2(\W^0)<Q_{j}^2(\W^k)<Q_{j}^2(\U)$,
$P_{j}^2(\U)+Q_{j}^2(\U)\ge c_{j}^2$.
Since $\lim Q_{j}^2(\W^k)=Q_{j}^2(\U)$ and
$\lim P_{j}^2(\W^k)=P_{j}^2(\U)$,
if we take $k^0=0$, then we can 
pick a large enough $k^1$ such that
\begin{itemize}
\item $P_{j}^2(\W^{k^0})>P_{j}^2(\W^{k^1})$,
and 
\item $Q_{j}^2(\W^{k^0})<Q_{j}^2(\W^{k^1})$,
 and
\item 
$P_{j}^2(\W^{k^0})+Q_{j}^2(\W^{k^1})>c_{j}^2$
(i.e., $(\W^{k^0},\W^{k^1})$ satisfies the inequality).
\end{itemize}
Similarly,
we can 
%Since $P_{j}^2(\W^{k^1})>P_{j}^2(\U)$
%and $\lim P_{j}^2(\W^k)=P_{j}^2(\U)$
%and $\lim Q_{j}^2(\W^k)=Q_{j}^2(\U)$,
%we can always 
pick a large enough $k^2>k^1$
 such that
\begin{itemize}
\item $P_{j}^2(\W^{k^1})>P_{j}^2(\W^{k^2})$, and
\item $Q_{j}^2(\W^{k^1})<Q_{j}^2(\W^{k^2})$, and 
\item  
$P_{j}^2(\W^{k^1})+Q_{j}^2(\W^{k^2})>c_{j}^2$
(i.e., $(\W^{k^1},\W^{k^2})$ satisfies the inequality).
\end{itemize}
It can be checked that $(\W^{k^0},\W^{k^2})$ also
satisfies the inequality.
This process can go on and, as a result, we obtain an infinite
 sequence
$$\W^{k^0},\cdots, \W^{k^i},\cdots$$
which satisfies:
\begin{itemize}
\item $P_{j}^2$ is in mode $\M(P_{j}^2)=\searrow$ on the sequence,
\item $Q_{j}^2$  is in mode $\M(Q_{j}^2)=\nearrow$ on the sequence,
\item $(\W^{k^{i_1}},\W^{k^{i_2}})$ 
satisfies the linear inequality
for all $i_1$ and $i_2$.
\end{itemize}
Therefore, the sequence
(as well as any subsequence)
 is a
 strong  $\omega$-chain of the linear inequality.
}

Thus, $S$ has a monotonic (of mode $\M$) and strong
 $\omega$-chain iff
formula (\ref{eq10}), which is definable in the additive theory
of reals, is satisfied by
some  $\U\in [0,1]^m$.
Hence,

\begin{lemma}\label{l3}
Let $S$ be  a conjunction of
$l$
dense linear equations
$P_{j}^1(\X)+Q_{j}^1(\X')=c_{j}^1$
and
$l$ dense linear inequalities
$P_{j}^2(\X)+Q_{j}^2(\X')>c_{j}^2$
defined in (\ref{eq1},\ref{eq2}).
Let $\M$ be a mode vector
on $\X$, $P_{j}^1, Q_{j}^1, P_{j}^2, Q_{j}^2$,
$1\le j\le l$.
Then, it is decidable
 whether
$S$ has a monotonic and strong 
 $\omega$-chain.
\end{lemma}

\section{The Existence of $\omega$-chains for
Discrete  Linear  Inequalities}\label{s5}

Assume that $T$ is a
conjunction of
$l$
discrete
linear inequalities
$P_{j}(\Y)+Q_{j}(\Y')>c_{j}$.
Let $\M$ be
a mode vector (on each integer  variable, each term $P_{j}$,
$Q_{j}$, $1\le j\le l$).
We use ``$\nearrow$",
``$\to$" and
``$\searrow$" to stand for
``unbounded increasing", ``flat"
and ``unbounded decreasing" modes, respectively.
Assume
that
$$\V^0,\cdots,\V^k,\cdots$$
is a monotonic and strong
$\omega$-chain
$\V^\omega$ of $T$.
Therefore,
\begin{equation}\label{eq11}
{\rm ~for~ any~} k_1<k_2,
T(\V^{k_1},\V^{k_2}).
\end{equation}
(\ref{eq11}) implies that,
for each $1\le j\le l$,
the mode $\M(P_{j})$ and the mode
$\M(Q_{j})$ only
have the following five combinations
(all the others
 are not possible):
\begin{itemize}
\item $\M(P_{j})=\nearrow$
and $\M(Q_{j})=\nearrow$,
\item $\M(P_{j})=\to$
and 
$\M(Q_{j})=\nearrow$,
\item $\M(P_{j})=\searrow$
and
$\M(Q_{j})=\nearrow$,
\item $\M(P_{j})=\nearrow$
and
$\M(Q_{j})=\to$,
\item $\M(P_{j})=\to$
and
$\M(Q_{j})=\to$.
\end{itemize}
If $\M(P_j)=\to$ (resp. $\M(Q_{j})=\to$),
we use $p_j$ (resp. $q_j$)
to stands for $P_j(\V^0)$ (resp. $Q_j(\V^0)$).
Similarly,
if $\M(y)=\to$, we use
$v_y$ to denote the component of
$y$ in $\V^0$.
Suppose $1\le j_1\ne j_2\le l$,
$\M(P_{j_1})=\searrow$
and $\M(Q_{j_1})=\nearrow$,
$\M(P_{j_2})=\nearrow$
and $\M(Q_{j_2})=\to$.
That is,
$\lim P_{j_1}(\V^k)=-\infty$,
$\lim Q_{j_1}(\V^k)=+\infty$,
$\lim P_{j_2}(\V^k)=+\infty$,
and for all $k$, 
$Q_{j_2}(\V^k)=q_{j_2}$.
 From (\ref{eq11}),
for all $k\ge 0$, we can pick
$\V^{k_1}$ and $\V^{k_2}$ such that
$T(\V^{k_1},\V^{k_2})$, and
\begin{itemize}
\item
$-k>P_{j_1}(\V^{k_1})>P_{j_1}(\V^{k_2})$, and
\item
$k<Q_{j_1}(\V^{k_1})<Q_{j_1}(\V^{k_2})$,
\end{itemize}
and
\begin{itemize}
\item
$k<P_{j_2}(\V^{k_1})<P_{j_2}(\V^{k_2})$, and
\item
$Q_{j_2}(\V^{k_1})=Q_{j_2}(\V^{k_2})=q_{j_2}$.
\end{itemize}
Similar statement can be made for all the valid
choices of
$\M(P_{j})$
and $\M(Q_{j})$, $1\le j\le l$, as well as
for $\M(y)$, $y\in \Y$.
That is,
for all $k\ge 0$, there are $\V^{k_1}$ and $\V^{k_2}$
such that
\begin{itemize}
\item $T(\V^{k_1},\V^{k_2})$,
\item $\V^{k_1}$ and $\V^{k_2}$ are consistent
with mode $\M(y)$ for each $y\in \Y$.
That is, for all $y\in \Y$,
 $\V^{k_1}(y)<\V^{k_2}(y)$ (resp. $=$, $>$)
and $k<\V^{k_1}(y)$ (resp.
$v_y=\V^{k_1}(y)$, $-k>\V^{k_1}(y)$)
if $\M(y)=\nearrow$ (resp. $\to, \searrow$),
where
$\V^{k_1}(y)$ is the component for  $y$ in vector $\V^{k_1}$.
\item For each $1\le j\le l$,
one of the following items  holds:
\begin{itemize}
\item 
$\M(P_{j})=\nearrow$
and $\M(Q_{j})=\nearrow$.
In this case,
$k<P_{j}(\V^{k_1})<P_{j}(\V^{k_2})$
and
$k<Q_{j}(\V^{k_1})<Q_{j}(\V^{k_2})$.
\item $\M(P_{j})=\to$
and
$\M(Q_{j})=\nearrow$. In this case,
$P_{j}(\V^{k_1})=P_{j}(\V^{k_2})=p_{j}$ and
$k<Q_{j}(\V^{k_1})<Q_{j}(\V^{k_2})$.
\item $\M(P_{j})=\searrow$
and
$\M(Q_{j})=\nearrow$.
In this case,
$-k>P_{j}(\V^{k_1})>P_{j}(\V^{k_2})$ and
$k<Q_{j}(\V^{k_1})<Q_{j}(\V^{k_2})$.
\item $\M(P_{j})=\nearrow$
and
$\M(Q_{j})=\to$.
In this case,
$k<P_{j}(\V^{k_1})<P_{j}(\V^{k_2})$ and
$Q_{j}(\V^{k_1})=Q_{j}(\V^{k_2})=q_{j}$.
\item $\M(P_{j})=\to$
and
$\M(Q_{j})=\to$.
In this case,
$P_{j}(\V^{k_1})=P_{j}(\V^{k_2})=p_{j}$
 and
$Q_{j}(\V^{k_1})=Q_{j}(\V^{k_2})=q_{j}$.
\end{itemize}
\end{itemize}
The above statement (replacing 
$\V^{k_1}$ with $\V$ and $\V^{k_2}$ with $\V'$)
can be written as
\begin{equation}\label{eq12}
\forall k\exists \V\exists \V' ~G(k, \vec C, \V,\V',\M)
\end{equation}
where $\vec C$ represents the tuple of all the constant values
$p_{j}$ and $q_{j}$, $1\le j\le l$, and $v_y$, $y\in \Y$.
Clearly, $G$ is a Presburger formula.
Conversely, we can show  the following lemma.

\begin{lemma}
If there are
 a $\vec C$ and a mode vector
$\M$ satisfying
(\ref{eq12}), then $T$ has a monotonic and
strong
 $\omega$-chain.
\end{lemma}

\proof{
Assume (\ref{eq12}) holds for some
$\vec C$ and a mode vector
$\M$. For $k=0$, according to
(\ref{eq12}), we pick
 $\V_0,\V_0'$ satisfying
$G(0, \vec C, \V_0,\V_0',\M)$. 
Take $$k=\max_{1\le j\le l}\{|P_j(\V_0')|,|Q_j(\V_0')|\}.$$
For this $k$, according to
(\ref{eq12}), we pick 
any $\V_1,\V_1'$ satisfying
$G(k,\vec C, \V_1,\V_1',\M)$.
What is the relationship among
$\V_0,\V_0',\V_1,\V_1'$?
Clearly, $T(\V_0,\V_0')$ and $T(\V_1,\V_1')$ hold.
More importantly,
$T(\V_0,\V_1)$ must be true.
This can be concluded from
the definition
of $G$ and the choice of $k$ and 
$\V_1$.
We can continue the procedure by 
taking 
$$k=\max_{1\le j\le l}\{|P_j(\V_1')|,|Q_j(\V_1')|\},$$
picking $\V_2,\V_2'$ from (\ref{eq12})
according to this $k$, and
concluding $T(\V_1,\V_2)$, etc.
Finally, we obtain an $\omega$-chain $\V_0,\cdots,\V_k,\cdots$ of $T$.
It is straightforward to verify that
the chain is monotonic (of mode $\M$) and strong.
}

In summary, for any $\M$,
$T$ has a
monotonic 
 and strong $\omega$-chain iff
\begin{equation}\label{eq4354}
\exists \vec C
\forall k\exists \V\exists \V' ~G(k, \vec C, \V,\V',\M).
\end{equation}
Since $G$ is Presburger, we have,

\begin{lemma}\label{l4}
Assume that $T$ is a
conjunction of
$l$
discrete
linear inequalities
$P_{j}(\Y)+Q_{j}(\Y')>c_{j}$.
Let $\M$ be a mode vector on
$\Y$, $P_{j}$ and
$Q_{j}$, $1\le j\le l$.
It is decidable whether $T$
has 
a monotonic and strong $\omega$-chain.
\end{lemma}

Now, we are ready to put 
Theorem \ref{Weis}, Lemma \ref{l1},
Lemma \ref{l2}, 
Lemma \ref{l3},  
Lemma \ref{l4}
 together and
conclude the main theorem.

\begin{theorem}\label{main}
It is decidable whether a transitive 
mixed linear relation
has an $\omega$-chain.
\end{theorem}

An upper bound for the time complexity of
the decidable result in Theorem \ref{main}
can be obtained as follows.
Let $R$ be given in (\ref{eq00}) whose length is $L$.
One can show that
the length of formula (\ref{eq10}) as
well as formula (\ref{eq4354})
is $O(L)$ (for any fixed choice of ${\cal M}$).
Using the complexity result given in 
\cite{Weis99},
the satisfiability of (\ref{eq10}) and
the satisfiability of 
(\ref{eq4354}) are decidable in time 
$2^{L^{(m+n)^{O(1)}}}$, for each fixed ${\cal M}$.
But since there are only (at most) 
$3^m3^n3^{3pl}3^{3pl}$  choices for ${\cal M}$,
whether $R$ has an $\omega$-chain is still
decidable in time $2^{L^{(m+n)^{O(1)}}}$.

Notice that the transitivity in Theorem \ref{main} is critical.
The existence of
an $\omega$-chain is undecidable for 
mixed linear relations.
The undecidability remains even for Presburger relations.
This is because a Presburger relation
can be used to encode one-step transitions
of a deterministic two-counter machine.
The negation of
the  halting problem (which is undecidable)
 for the machine
can be reduced to the existence of 
an $\omega$-chain for the Presburger relation.

\section{Applications}\label{s6}

In this section,
we will study various verification problems for
restricted infinite state systems containing
both dense counters and discrete counters.
We start with a general model.

\subsection{Mixed linear counter systems}

Let $M$ be a machine that is equipped with a number of
dense counters $\X$ and discrete counters $\Y$
and whose transitions involve changing control states
while changing counter values.
%OO Changed above 2 lines
A configuration of $M$ is a 
tuple consisting of a control state and counter values.
%% Changed above
Formally, $M$ is a tuple
$\langle S, \X, \Y, t\rangle$
where $t$ is the one-step transition such that
for each $s,s'\in S$, $t(s, \X, \Y, s', \X', \Y')$
indicates that $M$ transits
from a configuration $(s, \X, \Y)$
at
$s$ to another configuration $(s', \X', \Y')$
at $s'$.
$(s',\U',\V')$ is {\em reachable} from
$(s,\U,\V)$, written
$\T(s,\U,\V,s',\U',\V')$, if
there are $k$ (for some $k$)
 configurations
%OO Changed above
%%DD  add for some k
$(s_0,\U^0,\V^0),\cdots,(s_k,\U^k,\V^k)$ such that
$(s_0,\U^0,\V^0)=(s,\U,\V)$, 
$(s_k,\U^k,\V^k)=(s',\U',\V')$,
and $t(s_i,\U^i,\V^i,$ $s_{i+1},\U^{i+1},\V^{i+1})$ for all
$0\le i<k$. In this case, we say that
$(s,\U,\V)$ reaches $(s',\U',\V')$ through 
configurations $(s_i,\U^i,\V^i)$, $0\le i\le k$. 
Notice that
$\T$, called the {\em binary reachability} of $M$,
is the transitive closure of $t$.
$M$ is a {\em mixed linear counter system}
if, when $s$ and $s'$ are understood as bounded integer variables,
\begin{itemize}
\item $t(s, \X, \Y, s', \X', \Y')$ is a mixed linear relation,
\item $\T(s, \X, \Y, s', \X', \Y')$ is an (obviously transitive)
mixed linear relation.
\end{itemize}
%%DD changed above

Now, we assume that $M$ is a mixed linear counter system.
Let $I$ and $P$ be two subsets of configurations of $M$
both of which
are definable by mixed formulas.
There are two kinds of verification problems
we will consider.
$M$ is {\em $P$-safe from $I$} if
no configuration in $I$ reaches
a configuration in $P$.
The {\em mixed linear safety
problem} for $M$
is to decide
whether $M$ is $P$-safe from $I$.
An infinite sequence of configurations
$$(s_0,\U^0,\V^0),\cdots,(s_k,\U^k,\V^k),\cdots$$
of $M$ is {\em $P$-live from $I$} if
the following items hold:
\begin{itemize}
\item $(s_0,\U^0,\V^0)\in I$,
\item there are infinitely many $k$ such that $(s_k,\U^k,\V^k)\in P$,
and
\item for all $k\ge 0$,
$t(s_k,\U^k,\V^k,s_{k+1},\U^{k+1},\V^{k+1})$.
That is,
the sequence is an infinite execution of $M$.
\end{itemize}
$M$ is {\em $P$-live from $I$} if
there is an infinite sequence of configurations
that is $P$-live from $I$.
The {\em mixed linear liveness problem} for $M$
is to decide
whether $M$ is $P$-live from $I$.

These two problems can be further generalized.
Let $I$, $P_1,\cdots,P_k$ be
subsets of configurations of $M$
definable in
mixed formulas.
The $k$-mixed linear safety 
problem for $M$
is to decide whether
no configuration in 
$I$ reaches a configuration in $P_k$
through some configurations $c_1,...,c_{k-1}$
in $P_1,...,P_{k-1}$ respectively.
The $k$-mixed linear liveness
 problem for $M$
is to decide whether
there is an infinite execution of $M$
that is
$P_i$-live from $I$ for each $1\le i\le k$.
The $k$-mixed linear safety (resp. liveness)
problem is exactly
the mixed linear safety (resp. liveness) problem, when $k=1$.

\begin{theorem}\label{basic11}
(1). The $k$-mixed linear safety
problem for 
mixed linear counter systems
is
decidable for each $k$.
(2). The $k$-mixed linear liveness
problem for
mixed linear counter systems
is
decidable for each $k$.
\end{theorem}

\proof{
Let $M$ be a 
mixed linear counter system
with states $S$ and one-step transition $t$,
$I$ and $P_1,\cdots,P_k$ be sets (definable by mixed formulas)
 of configurations of $M$.
The proof of (1) is straightforward, since one can show that
the set of configurations
$c_0$ satisfying:
\begin{itemize}
\item $c_0$ in $I$,
\item there are configurations $c_1\in P_1,...,c_{k}\in P_k$ such that
$c_0$ reaches $c_k$ through $c_1$,...,$c_{k-1}$; i.e.,
$\T(c_0, c_1)$,...,$\T(c_{k-1},c_k)$,
\end{itemize}
%%DD changed above use c_0 for (s,X,Y)
is definable in a mixed formula (its satisfiability is decidable). 
 Now, we look at (2).
Define a formula $\hat \T$ as follows.
$\hat \T(s, \X, \Y,  s', \X', \Y')$ is true
iff there are
configurations
$(s_1, \X^1,\Y^1),\cdots,
(s_k,\X^k,\Y^k)$ such that,
\begin{itemize}
\item $(s, \X, \Y)$ is reachable from some configuration in $I$,
\item $(s_i,\X^i,\Y^i)$ satisfies $P_i$, for each $1\le i\le k$,
\item $(s, \X, \Y)$ reaches $(s_1, \X^1,\Y^1)$
(i.e., $\T(s, \X, \Y,s_1, \X^1,\Y^1)$),
\item $(s_i,\X^i,\Y^i)$ reaches $(s_{i+1},\X^{i+1},\Y^{i+1})$, for
each $1\le i<k$,
\item $(s_k,\X^k,\Y^k)$ reaches $(s', \X', \Y')$.
\end{itemize}
Since $M$ is a mixed linear counter system,
it is not hard to see that
$\hat \T$ is a transitive mixed linear relation.
(2) follows from Theorem \ref{main}, noticing that
$\hat \T$ has an $\omega$-chain iff
there is an infinite execution of $M$ that is
$P_i$-live from $I$ for each $1\le i\le k$.
}%end of proof

Consider the {\em eventuality} problem:
is there an infinite execution  of $M$
that starts from
some configuration in $I$ such that
$P$ is satisfied somewhere on the execution?
The problem is a special case of the mixed linear liveness
problem.
To see this, let $I'$ be the set of
configurations that are reachable from $I$ and satisfy $P$.
Obviously, the eventuality problem is equivalent to the problem whether
$M$
 is {\em true}-live from $I'$, which is decidable
({\em true} stands for the set of all configurations)
from Theorem \ref{basic11}.
We can modify the eventuality problem as follows:
is there an infinite execution  of $M$
that starts from
some configuration in $I$ such that
$P$ is satisfied by each configuration on the execution?
Unfortunately, this modified problem is undecidable for 
$M$, even when $M$ is a discrete timed automaton (cf. \cite{DSK01}
for a proof).
 
In practice, there are many counter models that have been found
being mixed linear. Applying Theorem \ref{basic11}
on these systems gives a number of new
%%DD change above
decidability results concerning safety/liveness verification.
 We first recall some definitions.

A {\em timed automaton} $\A$
is
a tuple
$$\langle S, \{x_1,\cdots,x_m\}, \C, Inv, R, C\rangle,$$
where
\begin{itemize}
\item $S$ is a finite set of {\em (control) states},
\item $x_1,\cdots,x_m$ are (dense) clocks,
\item $\C$ is the set of all
clock constraints over clocks $x_1,\cdots,x_m$; i.e.,
boolean combinations of formulas in the form of
$x_i-x_j\sim d$ or $x_i\sim d$ where $d$ is an integer,
$\sim$ stands for $<,>,\le,\ge,=$.
\item $Inv: S\to \C$ assigns a clock constraint
over  clocks
$x_1,\cdots,x_m$, called
an {\em invariant}, to each state,
\item $R: S\times S\to 2^{\{x_1,\cdots,x_m\}}$
assigns a subset of clocks
to a directed edge in $S\times S$,
\item $C: S\times S\to \C$
assigns a clock constraint over  clocks
$x_1,\cdots,x_m$,  called
a {\em reset condition},
 to a directed edge in $S\times S$.
\end{itemize}
The semantics of $\A$ is defined as follows.
A configuration $(s,\U)$  is a pair of
a control state $s$
and
a tuple $\U$ of clock values.
A transition
is either a progress transition or
a reset transition.
A
progress transition makes all the clocks
 synchronously progress
by a positive amount, during which the invariant is
consistently satisfied, while the automaton 
remains at the same control state.
A reset transition, by moving
 from state $s_1$ to state $s_2$,
 resets
 every clock in $R(s_1,s_2)$ to 0 and keeps all the other clocks unchanged.
In addition,
clock values before the transition satisfy
the invariant $Inv(s_1)$ and the reset condition $C(s_1,s_2)$;
clock values after the transition satisfy
the invariant $Inv(s_2)$.
In particular,
when the clocks are integer-valued
(and hence clocks are incremented by some positive integral amount
in a progress transition),
$\A$ is called a
{\em discrete timed automaton}.
The following characterization has recently been established
\cite{CJ99}.
%%DD change above references
\begin{theorem}\label{cj}
Timed automata, as well as
discrete timed automata,
are mixed linear counter systems.
\end{theorem}
Hence, from Theorem \ref{basic11}, the following corollary is obtained.
\begin{corollary}\label{b11}

(1). The $k$-mixed linear safety problem is decidable for
timed automata as well as for discrete timed automata
\cite{CJ99}.
%%DD change above references

(2). The $k$-mixed linear liveness problem is decidable for
discrete timed automata \cite{DSK01}.

(3). The $k$-mixed linear liveness problem is decidable for
timed automata.
\end{corollary}

A (free) counter is an integer variable that can be
tested against 0, incremented by 1, decremented by 1, and
stay unchanged.
A timed automaton can be augmented with
counters by integrating a reset transition with
a counter operation. A counter in a timed automaton
is {\em reversal-bounded} if
there is a number $r$ such that, during
any execution of
the automaton,
the counter  changes mode between
nondecreasing and nonincreasing
for at most $r$ times.
Let $\A$ be a timed automaton augmented with
a finite number of reversal-bounded counters and
one free counter.
Now, a configuration $(s,\U, \V)$ of $\A$ is
a tuple of
a control state $s$, dense clock values $\U$ and
counter values $\V$.
When $\A$ does not contain any clocks, it is 
a finite automaton augmented with
reversal-bounded counters and one free counter.
\begin{theorem}\label{cj1}

(1). Discrete
timed automata augmented with
reversal-bounded counters and one free counter
are mixed linear counter systems \cite{DIBKS00}. 

(2). Timed automata augmented with
reversal-bounded counters and one free counter
are mixed linear counter systems \cite{D01}.
\end{theorem}
Hence, from Theorem \ref{basic11}, the following corollary is obtained.
\begin{corollary}\label{b12}

(1). The $k$-mixed linear safety problem is decidable for
 discrete timed automata 
augmented with
reversal-bounded counters and one free counter \cite{DIBKS00}.

(2). The $k$-mixed linear safety problem is decidable for
timed automata
augmented with
reversal-bounded counters and one free counter \cite{D01}.

(3). The $k$-mixed linear liveness problem is decidable for
finite automata augmented with
reversal-bounded counters and one free counter \cite{DIS01}.

(4). The $k$-mixed linear liveness problem is decidable for
timed automata (as well as
discrete timed automata)
augmented with
reversal-bounded counters and one free counter.
\end{corollary}

Corollary \ref{b11} (3) and
Corollary \ref{b12} (4) are new decidability results.
One shall notice that the loop analysis techniques
presented in \cite{DSK01,DIS01} to show Corollary \ref{b11} (2) and
Corollary \ref{b12} (3) can not be easily used to prove 
our new results.
The corollaries
can be used to automatically verify
a class of non-region safety and
liveness properties
that, previously, could not be done using the traditional
region technique \cite{AD94}.
Below, we look at an example of liveness verification.
Consider a system $S$
of two concurrent
processes $S_1$ and $S_2$.
The two processes may use
a counting semaphore
to perform
 concurrency control.
In some applications,
we would like to ensure that
the concurrency control makes $S$ starvation-free;
i.e.,
it is not possible that the composite system $S$, starting from 
some initial configuration,
%%DD change above
executes for some finite number of steps and then
$S_1$ solely executes forever (in this case, $S_2$ starves).
We use $S'$ to denote the system
that behaves like $S$ then, nondeterministically,
behaves like $S_1$ afterwards.
It is observed that $S_2$ starves iff
$S'$ has an $\omega$-chain (i.e.,
$S'$ is $true$-live from the initial configuration).
Now, we suppose 
that $S_1$ and $S_2$ are 
real-time processes modeled as discrete
timed automata. A free counter is used for
the counting semaphore.
 From Corollary \ref{b12} (4),
whether $S_2$ starves can be automatically verified.

Besides mixed linear safety/liveness problems, one may also be interested in
a class of boundedness problems as below. Let $M$ be a mixed linear counter 
system with dense counters $\X$ and discrete counters $\Y$.
Let $I$ be a set
of configurations definable in a 
mixed formula.
We use $l$ to denote a linear combination of 
$\X$ and $\Y$; i.e.,
$l=\Sigma a_ix_i + \Sigma b_jy_j +c$ with $a_i,b_j, c$ integers.
Let $l_1,...,l_p$ be $p$ such linear combinations. 
Are there numbers $B_1,...,B_p$ such that,
starting from a configuration in $I$,
$M$ can only reach a configuration
satisfying $l_i\le B_i$ for each $1\le i\le p$?
This boundedness problem can be easily shown decidable,
since the question is equivalent to the satisfiability
(for $B_1,...,B_p$) of the
following mixed formula:
$\forall \alpha,\beta: 
\alpha\in I\land \T(\alpha,\beta)\to $ ``$\beta$ satisfies
$l_i\le B_i$ for each $1\le i\le p$".
One may also ask a slightly different question:
\begin{quote}
(*) For each infinite execution starting from $I$,
are there $p\ge 1$ numbers $B_1,...,B_p$ such that
every configuration on the execution satisfies
$l_i\le B_i$ for each $1\le i\le p$?
\end{quote}
We call this question as the {\em mixed linear boundedness
problem}, whose decidability is not obvious.
\begin{theorem}\label{bound}
The mixed linear boundedness
problem is decidable for mixed linear
counter systems.
\end{theorem}

\proof{
Let $M$ be a mixed linear
counter system. Without loss of generality, we assume $p=1$
(the other cases for $p$ are similar).
That is, we are given one linear combination $l$.
An infinite execution is {\em unbounded} for $l$ if
for any $B$ there is some configuration on the execution 
satisfying $l>B$.
It suffices for us to consider the negation of 
the question statement (*):
whether there is an unbounded
infinite execution starting from $I$.
The proof uses the idea of Theorem \ref{basic11}.
Define a formula $\hat \T$ as follows.
$\hat \T(s, \X, \Y,  s', \X', \Y')$ is true
iff the following two items are true:
\begin{itemize}
\item $(s, \X, \Y)$ is reachable from some configuration in $I$,
\item $(s, \X, \Y)$ reaches $(s', \X', \Y')$; i.e.,
$\T(s, \X, \Y,  s', \X', \Y')$,
\item $l(\X, \Y)+1 \le l(\X', \Y')$.
\end{itemize}
The result follows immediately,
noticing that $\hat \T$ is a transitive mixed linear relation and
$\hat \T$ has an $\omega$-chain iff
$M$ has an
unbounded
infinite execution from $I$.
}%end of proof

 From their
 proofs,
Theorem \ref{bound} and Theorem \ref{basic11} (2)
can be combined.
For instance, the following question is decidable:
is there an infinite execution of $M$ that is $P$-live from $I$ and
that is unbounded for $l$?

Notice that, in (*), the bounds $B_1,...,B_p$ are not uniform over all
the infinite executions. To make them
uniform, one might ask another different question by switching the 
quantifications in (*): 
\begin{quote}
(**) Are there numbers $B_1,...,B_p$ such that,
for each infinite execution starting from $I$,
every configuration on the execution satisfies
$l_i\le B_i$ for each $1\le i\le p$?
\end{quote}
Currently, we do not know whether (**) is decidable or not.
We leave this as an open question. However,
the following question (by making $B_1,...,B_p$ in (**) fixed, e.g., 0)
\begin{quote}
is it true that,
for each infinite execution starting from $I$,
every configuration on the execution satisfies
$l_i\le 0$ for each $1\le i\le p$?
\end{quote}
is decidable, since its negation is equivalent to 
an eventuality problem.

One can easily find applications for Theorem \ref{bound}.
For instance, consider a system with
two concurrent real-time 
processes
running on one CPU. The processes
are modeled as two discrete timed automata using
a lock semaphore to achieve concurrency and using
clocks to enforce timing constraints.
The system is designed to be non-terminating
and some fairness
constraints are expected. 
We use $t_1$ (resp. $t_2$) to denote the total time that process 1 (resp. 
process 2)
takes the CPU so far.
One such constraint could be as follows.
There is no infinite execution of the system on which
the difference $|t_1-t_2|$
is unbounded.
This constraint can be automatically verified due to Theorem
\ref{bound} and the fact, from Theorem \ref{cj1},
 that the system,
a
discrete timed automaton augmented with two monotonic
(and hence reversal-bounded)
counters $t_1$ and $t_2$,
is a mixed linear counter system.

\subsection{Timed pushdown systems}

There has been much interesting work 
on various verification problems for pushdown systems
\cite{BR00,BEM97,BER95,D01,DIBKS00,DIS01,ES01,FWW97}.
Studying pushdown systems is important,
since
they are directly related to recursive programs and processes.
In this subsection, we will study 
pushdown systems with
discrete clocks and reversal-bounded counters.
Safety verification for these systems
is discussed in \cite{DIBKS00}.
Here, we investigate the
mixed linear liveness 
 problem (since now we have only discrete variables, we call the  
problem as the Presburger liveness problem).

\renewcommand{\#}{{\bf n}}

As we mentioned before,
a  timed automaton can be augmented with reversal-bounded counters.
Here we only consider discrete clocks that take integer values.
The discrete timed automaton can be further augmented with
a pushdown stack.
The  resulting machine $\A$ is called
a discrete pushdown
timed automaton with reversal-bounded counters.
In addition to counter operations and clock operations,
$\A$  can push a symbol on the top of the stack,
pop the top symbol from the stack, and
test whether the top symbol of the stack equals some symbol.
A configuration of $\A$ is a tuple
of
a control state, discrete clock values, counter values,
and a stack word.
The binary reachability $\T$
is the set of configurations pairs such that one can reach the other
in $\A$.
Each stack word $w$ 
corresponds to an integer tuple
$\#=(\#_{a^1},\cdots,\#_{a^l})$,
where $\{a^1,\cdots,a^l\}$ is the stack alphabet and
each count $\#_{a^i}$ stands for the number of symbol $a^i$ in $w$.
The tuple $\#$ is also called the {\em stack word counts} for $w$.
In this way,
a set $C$ of configurations corresponds to
a predicate on states, clock values, counter
values, and stack word counts.
$C$ is Presburger if the predicate is definable by a Presburger formula.
$C$ is {\em commutative} if,
for any configurations $c$ and $c'$
satisfying
that $c$ and $c'$ are the same except that
the stack word in $c$ is a permutation of
the stack word in $c'$,
$c\in C$ iff $c'\in C$.
In this case, the predicate exactly characterizes
the set $C$.
Let $I$ and $P$ be two Presburger subsets of configurations.
We say $\A$ is {\em $P$-live from $I$} if
there is
an infinite sequence $c^0,\cdots,c^k,\cdots$ such that
(1). $c^0\in I$, (2). 
for all $k\ge 0$, $\T(c^k,c^{k+1})$,
and (3). $c^k\in P$ for infinitely many $k$.
The Presburger liveness problem for $\A$ is whether
$\A$ is {\em $P$-live from $I$}, given
$I$ and $P$ two Presburger subsets of configurations.

\begin{theorem}\label{pio}
The Presburger liveness problem for
discrete pushdown
timed automata with reversal-bounded counters is
decidable.
\end{theorem}
\proof{
Let $\A$ be a discrete pushdown
timed automaton with reversal-bounded counters.
We use $\Y$ to denote the discrete clocks and
counters in $\A$.
We use $\#$ to denote an integer tuple of stack word counts.
Let $I$ and $P$ be two Presburger subsets of configurations of $\A$.
Define
$\hat \T$ as follows.
$\hat \T(s, \Y,  {\bf \#}, a, s',\Y',  {\bf \#}', a')$
is true iff
there are two stack words $w$ and $w'$ (called {\em witnesses})
such that 
\begin{itemize}
\item (Condition 1) $w$ is a (not necessarily proper) prefix of $w'$,
\item (Condition 2.1)
$w$ ends with stack symbol
$a$ (i.e., $a$ is the top symbol of the stack word $w$),
\item (Condition 2.2)
$w'$  ends with stack symbol
$a'$,
\item (Condition 3.1) ${\bf \#}$ is the stack word counts for $w$,
\item (Condition 3.2) ${\bf \#}'$ is the stack word counts for $w'$,
\item (Condition 4) configuration $(s, \Y,  w)$ is reachable 
from some configuration in
$I$,
\item (Condition 5) configuration $(s, \Y,  w)$ reaches
configuration $(s',\Y', w')$
through
 a sequence of 
moves in $\A$,
 during which the top symbol $a$ of $w$ 
is not popped out and during which there is a configuration in $P$.
\end{itemize}
Assume that $w''$ and $w'''$ witness 
$\hat \T(s, \Y,  {\bf \#}, a, s',\Y',  {\bf \#}', a')$.
Observe that, for any $w$
satisfying 
(Condition 2.1), (Condition 3.1) and (Condition 4),
$w$ and $w'=w+(w'''-w'')$
(i.e., $w$ concatenated with the result of deleting
the prefix $w''$ from $w'''$)
also
witness
$\hat \T(s, \Y,  {\bf \#}, a, s',\Y',  {\bf \#}', a')$.
 The reason is as follows.
According to (Condition 5),
the top $a$ of $w''$ will not be popped out.
That is, the content (instead of counts)
of $w''$ is insensitive to (Condition 5). Therefore,
(Condition 5) still holds
when $w''$ is replaced with $w$
as long as the prefix $w''$ of $w'''$ is also replaced with
$w$; i.e.,
(Condition 5) still holds
for $w$ and $w'$.
This observation will be used in proving the following claim.

\medskip

({\bf Claim 1}) $\hat \T$ has an $\omega$-chain iff
$\A$ is $P$-live from $I$.

\medskip

\noindent Proof of ({\bf Claim 1}).
($\Rightarrow$). 
Assume $\hat \T$ has an $\omega$-chain
$$(s_0,\V_0,\#_0,a_0),\cdots,
(s_k,\V_k,\#_k, a_k),\cdots.$$
Therefore, for each $k$,
 we have  a pair of stack words
$w_k$ and $w_k'$ that witness the fact of
$\hat \T(s_k,\V_k,\#_k,a_k,s_{k+1},\V_{k+1},\#_{k+1}, a_{k+1})$.
Now, take $w''_0=w_0$, and for all $k\ge 1$, 
$w_k''=w_0+(w_0'-w_0)+\cdots+(w_{k-1}'-w_{k-1})$.
Using the above observation, it can be easily shown that,
for any $k\ge 0$,
$w_k''$ and $w_{k+1}''$ witness
$$\hat \T(s_k,\V_k,\#_k,a_k,s_{k+1},\V_{k+1},\#_{k+1}, a_{k+1}).$$
Applying (Condition 4) on
configuration $(s_0,\V_0,w''_0)$ and
(Condition 5) on configurations 
$(s_k,\V_k,w_k'')$ and $(s_{k+1},\V_{k+1},w_{k+1}'')$
for all $k\ge 0$,
we can show $\A$ is $P$-live from $I$.

($\Leftarrow$).
Assume $\A$ is $P$-live from $I$.
That is, there is
an infinite sequence $c^0,\cdots,$ $c^k,\cdots$ such that
(1). $c^0\in I$, (2).
for all $k\ge 0$, $\T(c^k,c^{k+1})$,
and (3). $c^k\in P$ for infinitely many $k$.
Without loss of generality, we assume that 
$\A$ leads $c_k$ to $c_{k+1}$
by running exactly one move, for all $k\ge 0$.
Therefore, the stack word $w_k$
in $c_k$ and the stack word $w_{k+1}$ in
$c_{k+1}$ satisfy one of the following conditions:
(1). $w_k=w_{k+1}a$; i.e., the move pops $a$  for some symbol $a$,
(2). $w_{k+1}=w_ka$; i.e., the move pushes $a$  for some symbol $a$,
(3). $w_{k+1}=w_k$; i.e., the move does not change the stack.  
Notice that the stack has a special 
bottom symbol $Z_0$;
i.e., every $w_k$ starts with $Z_0$.
The following technique has been used in several places
(e.g., \cite{HJ79,BEM97}).
%%DD added references
For the sequence of the stack words
$w_0,
\cdots,
w_k,\cdots$, 
define a strictly increasing sequence $k_0,\cdots,k_i,\cdots$ as follows.

$k_0$ is picked such that
$w_{k_0}$ is a prefix of each $w_k$ with $k\ge 0$;

$k_1>k_0$ is picked such that
$w_{k_1}$ is a prefix of each $w_k$ with $k>k_0$;

$k_2>k_1$  is picked such that
$w_{k_2}$ is a prefix of each $w_k$ with $k>k_1$;
etc. 

\noindent Such a sequence always exists.
Clearly, each $w_{k_i}$ is a prefix of $w_{k_{i+1}}$ and
from configuration
$c_{k_i}$ to configuration $c_{k_{i+1}}$,
the top symbol of
$w_{k_i}$ is not popped out.
Since there are infinitely many $k$ with $c_k\in P$,
there is a strictly increasing sequence
$i^0,\cdots,i^j,\cdots$  such that, for all $j$,
there is a $k$ satisfying
$c_k\in P$ and $k_{i^j}<k<k_{i^{j+1}}$.
For each $j\ge 0$,
we use $(s_j,\V_j,\#_j,a_j)$ to denote
the control state, clock and counter values,
the count vector of the stack word, and the top symbol
of the stack word, respectively in
configuration $c_{k_{i^j}}$.
It is left to the reader to check
$$(s_0,\V_0,\#_0,a_0),\cdots,
(s_j,\V_j,\#_j,a_j),\cdots$$
is an $\omega$-chain of $\hat \T$, where, for all $j\ge 0$,
$\hat \T(s_j,\V_j,\#_j,a_j,s_{j+1},
\V_{j+1},\#_{j+1},a_{j+1})$ is witnessed by
$w_{k_{i^j}}$ and $w_{k_{i^{j+1}}}$.

Therefore, ({\bf Claim 1}) is proved. Next, we are going to show that,

\medskip

({\bf Claim 2}).
$\hat \T(s, \Y,  {\bf \#}, a, s',\Y',  {\bf \#}', a')$
is a Presburger formula (when $s,s',a,a'$ are understood as
bounded integer variables).

\medskip

\noindent Proof of ({\bf Claim 2}).
We build a machine $M$ that accepts the domain (which are integer
tuples) of $\hat \T$.
Then we argue that integer 
tuples accepted by $M$ are definable by a Presburger formula.
$M$ is a machine with a one-way input tape and a pushdown stack.
$M$ is also equipped with a number of counters, among which
each clock in $\A$ corresponds a {\em clock-counter} in $M$ and
each reversal-bounded counter in $\A$ corresponds
to a {\em rv-counter} in $M$.
In addition,
$M$ contains a count-counter for each stack symbol
and contains a number of other auxiliary counters.
Whenever
$M$ pushes $a$ to (resp. pops $a$ from)
the stack,
the count-counter for $a$ is incremented (resp. decremented)
by one.
So, a count-counter is
 used to record the
number of a stack symbol in a stack word.
$M$ works as follows.
Given an input
$$(s, \Y,  {\bf \#}, a, s',\Y',  {\bf \#}', a')$$
on $M$'s input tape, where each integer in the above tuple
is encoded as a unary string and separated by a delimiter,
$\M$ starts to simulate $\A$ as follows.
$M$ guesses a control state for $\A$,
a value for each clock-counter and
a value for each rv-counter.
At this moment, $M$ makes sure that the stack is empty and
each count-counter is 0.
Then $M$ guesses a stack word (by nondeterministically pushing
symbols)
and updates the count-counters accordingly.
At some moment,
$M$ decides that $I$ is satisfied by checking
that the guessed control state, the clock-counter values,
the rv-counter values, and the count-counters
satisfy $I$.
Doing this needs some auxiliary counters
and needs only a finite number of counter reversals,
since $I$ is Presburger \cite{I78}.
When this is checked out,
$M$ starts to simulate $\A$ (from the guessed state) using its own
stack for
the stack in $\A$,
its own clock-counters for the clocks in $\A$ and
its own
rv-counters for the reversal-bounded counters in $\A$.
All the transitions of $\A$ are faithfully simulated by 
$M$. In addition,
whenever
$\A$ pushes $a$ to (resp. pops $a$ from)
the stack,
$M$ increments (resp. decrements)
 the count-counter for $a$ by one.
Nondeterministically at some moment,
$M$ decides to read the input tape by suspending
the simulation. Then, $M$ makes sure that
the first half of the input
$(s,\Y,\#,a)$ are consistent
with the current configuration of 
$\A$.
That is, the control state of $\A$ (remembered in $\M$'s finite 
control) is $s$,
clock-counters and rv-counters have the same values
as in $\Y$ (doing this needs auxiliary reversal-bounded
counters),
the stack top symbol is $a$, and
count-counters have the same values as in $\#$
(doing this also needs auxiliary reversal-bounded
counters).
When these are checked out,
(Condition 2.1),
(Condition 3.1) and
(Condition 4) are satisfied for the current configuration
$(s,\Y,\#,a)$
of $\A$.

Then, $M$ replaces the stack top symbol $a$ with a new symbol $\hat a$
and
resumes the simulation of $\A$.
$M$ makes sure that
the simulation afterwards will not 
pop 
the new symbol out of $M$'s stack.
Nondeterministically at some moment later,
$M$ decides that the current
configuration of $\A$ satisfies
$P$. $M$ checks that this is indeed true using
its own counters.
Similar to the previous scenario for $I$,
this checking needs only a finite number of 
counter reversals and needs
other auxiliary reversal-bounded counters.
When this is checked out, $M$ resumes the simulation of $\A$.
Again,
nondeterministically
at some moment later, 
$M$ shuts down the simulation and 
compares the rest of the input tape $(s',\Y',\#',a')$
with the control state of $\A$ in $M$'s finite control,
the clock-counter and rv-counter values of $M$,
the count-counter values, and the top symbol of the stack.
The comparisons make sure that (Condition 1), 
(Condition 2.2), (Condition 3.2) and
(Condition 5) are satisfied by the 
current configuration of $\A$.
$M$ accepts the input if the comparisons are
 successful.
Clearly,
$M$ accepts exactly the domain of $\hat \T$.

What are the counters in $M$? they are
clock-counters, rv-counters, count-counters, and 
a number of other auxiliary reversal-bounded counters.
All of them  are reversal-bounded except
the clock-counters and the count-counters.
Each count-counter $n_a$ can be treated as 
the difference $n^+_a-n^-_a$ 
of two reversal-bounded counters $n^+_a$ and $n^-_a$:
$n^+_a$ (resp. $n^-_a$) is used to record the number of
pushes (resp. pops) of $a$. So, each count-counter
can be simulated by two reversal-bounded counters.
How about clock-counters?
In \cite{DIBKS00} (see also its full version),
a technique is proposed such that,
as far as binary reachability is concerned,
discrete clocks can be simulated by reversal-bounded counters
\footnote{More precisely,
discrete clocks in $\A$ can be replaced by reversal-bounded counters
(the result is called $\A'$)
such that, whenever $c_1$ can reach $c_2$ in $\A$,
$c_1$ can reach $c_2$ in $\A'$ \cite{DIBKS00}.
}. Therefore, clock-counters can be made reversal-bounded
from the start of simulating $\A$
to the moment checking $P$,
 and,
from the moment checking $P$ to shutting down $\A$.
Hence, $M$  only has
 reversal-bounded counters as well
as a pushdown stack. 
Therefore,
$M$ is a reversal-bounded multicounter machine
with a pushdown stack and a one-way input tape
(NPCM).
It is known that 
NPCMs accepts semilinear languages \cite{I78}.
In particular, since $M$  accepts
a language in the form of integer tuples,
the language is definable by a Presburger formula \cite{I78}.
Hence, $\hat \T$ is  Presburger.
Therefore, ({\bf Claim 2}) is proved.

Since a Presburger formula is a special form
of a mixed linear relation,
Theorem \ref{pio} is followed from ({\bf Claim 1}), ({\bf Claim 2}),
and Theorem \ref{main}.
}%end of proof

We are not able to extend the result of
Theorem \ref{pio} to dense clocks. 
The pattern technique \cite{D01}
that abstracts a dense clock into a
discrete clock and a pattern  does not
 apply here. This is because
the abstraction maintains the exact
binary reachability of dense clocks, but does not maintain
the exact dense clock values between the binary reachability.
Timed pushdown systems with reversal-bounded counters
 dealt in Theorem \ref{pio} also have a lot of applications.
For instance, it can be used to model
some real-time recursive concurrent programs.
The reversal-bounded counters can also be used to count
the number of external events -- these counts can be later used to
specify some fairness constraints on the environment.

\section{Conclusions}\label{s7}

In this paper, we showed that
%OO Changed above line
it is decidable whether a transitive 
mixed linear relation has an $\omega$-chain.
Using this main theorem,
we were able to establish, within a unified
%OO were above
framework,  a number of liveness
verification results on generalized timed automata.
More precisely, we proved that
%OO proved above
(1) the mixed linear liveness problem for
timed automata with dense clocks, reversal-bounded
counters, and a free counter is decidable, and
(2) the Presburger liveness problem for
timed automata with discrete clocks, reversal-bounded
counters, and a pushdown stack is decidable.
The results
can be used to 
analyze some fairness constraints
(e.g., livelock-free and starvation-free)
 for infinite-state concurrent systems.

Our results 
are useful in formulating a
decidable subset of linear temporal logic (LTL)
for a class of timed automata augmented with counters.
Let $\A$ be a timed automaton with dense clocks, reversal-bounded
counters, and a free counter.
The set of linear temporal logic formulas $\cal L_A$
with respect to $\A$
is defined by the following grammar:
$$\phi:= P | \neg \phi | \phi\land \phi | \bigcirc \phi | \phi U \phi$$
where $P$ is a  set of configurations of $\A$ definable by
a mixed formula (on control states, dense clocks,
reversal-bounded
counters, and the free counter).
$\bigcirc$ denotes ``next", and
$U$ denotes ``until".
Formulas in $\cal L_A$
are interpreted on infinite execution sequences $p$
 of configurations of $\A$
in the usual way. 
This logic is very similar to  
the Presburger LTL for timed
automata with discrete clocks \cite{DSK01}
except that $P$ is a mixed formula instead of a Presburger formula.

The {\em satisfiability-checking problem}
 is to 
check, given $\A$ and
$\phi\in \cal L_A$,
 whether there exists an infinite execution $p$ of
$\A$ with $p\models \phi$.
 From Corollary \ref{b12},
the satisfiability-checking 
problems are decidable for the following LTL formulas:
\begin{itemize}
\item $I\land \Box\Diamond P$. 
\item $I\land \Diamond P$. 
\item $I\land \Box\Diamond P\land \Box\Diamond Q$.
\end{itemize}
In our previous paper \cite{DSK01},
the first two items as above were shown
but only for timed automata with
discrete clocks. In the same paper,
the last item as above was left open.

Some work needs to be done in the future
in formulating
an exact decidable
subset (broader than the subset in Comon and Cortier \cite{CC00})
 of
$\cal L_A$
for 
satisfiability-checking.
Notice that the entire 
$\cal L_A$ is undecidable for
satisfiability-checking/model-checking,
even when the next operator is dropped from the logic.
This is because
the satisfiability-checking
problem for $\Box P$ is undecidable,
when $\A$ is a discrete timed automaton, as shown in \cite{DSK01}.

A similar decidable
subset of LTL formulas $\cal L_A$ could be formulated 
for discrete timed pushdown systems,
by combining Theorem \ref{pio},
the results in \cite{DIBKS00} and \cite{ID01}.
Another
 issue is
on the complexity analysis
 of the decision procedures
presented in Theorem \ref{basic11}
and Theorem \ref{pio}.
However, this issue is related to
the complexity for the emptiness problem of 
NPCMs, which is still unknown, though it is believed that
it can be derived along Gurari and Ibarra \cite{GI81}.

\nonumsection{Acknowledgements}
We would like to thank the following students at WSU
for reading an earlier draft of this paper:
K.  Gjermundrod,
H. He, A. Khodjanov, C. Li, J. Nelson,
G. Xie, and L. Yang.
The work by 
Oscar H. Ibarra
has been
  supported in part by NSF Grant IIS-0101134. 
Thanks also go to the anonymous referees for many useful suggestions.

\nonumsection{References}
%References are to be listed in the order cited in the text. Use the style
%shown in the following examples. For journal names, use the standard
%abbreviations. Typeset references in 9 pt Times Roman.

\end{document}